\documentstyle[prb,aps]{revtex} %twocolumn%
\begin{document}
\draft
\title{Incommensurate nodes in the energy spectrum of weakly coupled
antiferromagnetic Heisenberg ladders}
\author{M. Azzouz, B. Dumoulin}
\address{Centre de Recherche en Physique du Solide
et D\'epartement de Physique\\ Universit\'e de Sherbrooke, Sherbrooke
Qu\'ebec,
 J1K 2R1 Canada}
\author{A. Benyoussef}
\address{Laboratoire de Magn\'etisme et de Physique des Hautes \'Energies, et
D\'epartement de Physique\\ Universit\'e Mohammed V, B.P. 1014, Rabat, Maroc}
\date{\today}
\maketitle
\begin{abstract}
Heisenberg ladders are investigated using the bond-mean-field theory
[M.Azzouz, Phys.Rev.B {\bf 48}, 6136 (1993)].
The zero inter-ladder coupling energy gap, the uniform spin susceptibility and
the nuclear magnetic resonance
spin-relaxation rate are calculated as a function of temperature and magnetic field.
For weakly coupled ladders, the energy spectrum vanishes 
at incommensurate wavevectors giving rise to nodes.
As a consequence, the spin susceptibility becomes linear 
at low temperature.
Our results for the single ladder successfully compare 
to experiments on SrCu$_2$O$_3$ and (VO)$_2$P$_2$O$_7$ materials
and new predictions concerning the coupling to the magnetic field are made.
\end{abstract}
%
%\section{introduction}
%
Among materials that contributed to the increased interest in low-dimensional spin 
systems, one finds ladder compounds like vanadyl-pyrophosphate
(VO)$_2$P$_2$O$_7$\cite{Johnston} or Cu-O layers of 
SrCu$_2$O$_3$.\cite{hiroi,azuma}
These can be modeled by the Heisenberg Hamiltonian 
on a ladder geometry.
Here, the general Heisenberg model 
\begin{eqnarray}
H= J_\perp\sum_{i,j}{\bf S}_{i,2j}\cdot{\bf S}_{i,2j+1}
+ J'_\perp\sum_{i,j}{\bf S}_{i,2j+1}\cdot{\bf S}_{i,2j+2}
 + J\sum_{i,j}{\bf S}_{i,j}\cdot{\bf S}_{i+1,j} - h\sum_{i,j}S_{i,j}^z
\label{hamiltonian}
\end{eqnarray}
for an array of 
coupled ladders in a 
uniform magnetic field $h$ along the z-direction is considered.
The quantities $J$, $J_\perp$ and $J'_\perp$ stand respectively for intrachain, 
transverse intra-ladder and inter-ladder positive exchange coupling constants.
It is now well established that any finite transverse coupling opens an energy 
gap, $E_g$, in elementary spin excitations for the single 
ladder $(J_\perp'=0)$.\cite{azzouz1,preprint,barnes1,dagotto}
As a consequence of this energy gap,
thermodynamical quantities like the uniform susceptibilty, $\chi'$,
and the nuclear magnetic resonance (NMR) spin-relaxation rate, 
$T_1^{-1}$, show an activated behavior
at low temperatures,\cite{troyer} $T<E_g$. 
Recently an important numerical work has been reported by Sandvik {\it et al.} 
\cite{sandvik} who used Quantum Monte Carlo (QMC) technique
to calculate the magnetic susceptibility and the NMR relaxation rate as
a function of temperature for the SrCu$_2$O$_3$ material. However, comparisons
to experiments did not lead to a clear conclusion concerning the magnitude
of the longitudinal, $J$, and transverse, $J_\perp$, exchange couplings. 
The best fits to Azuma {\it et al.}\cite{azuma} results 
they reported for $\chi'$ and $T_1^{-1}$ were 
obtained using different values, namely
$J=J_\perp=850$ K for $\chi'$ and $J=J_\perp=1200$ K for $1/T_1$.
It is then interesting to address the question of the relevance of the 
Heisenberg ladder for this system.
Frischmuth {\it et al.}\cite{frischmuth} numerically
studied (also using QMC method) the temperature dependence of the uniform 
susceptibility for Heisenberg ladders with up to 6 legs.
They came to the conclusion that the gap rapidly decreases when
the number of legs increases (for even number of legs).

The Hamiltonian (\ref{hamiltonian}) 
is studied by the bond-mean-field theory (BMFT)
explained in detail in Refs.[\cite{azzouz1,azzouz3,azzouz2}].
In this approach,
(\ref{hamiltonian}) becomes a spinless fermion Hamiltonian
\begin{eqnarray}
H=&&{J\over2}\sum_{i,j}c^{\dag}_{i+1,j} c_{i,j}e^{-i\Phi}
+{J_\perp\over2}\sum_{i,2j}c^{\dag}_{i,2j+1} c_{i,2j}e^{-i\Phi_\perp}
+{J'_\perp\over2}\sum_{i,2j}c^{\dag}_{i,2j+2} c_{i,2j+1}e^{-i\Phi'_\perp}
+J\sum_{i,j}(n_{i,j}-{1\over2})(n_{i+1,j}-{1\over2}) \cr
&& + J_{\perp}\sum_{i,j}
(n_{i,2j}-{1\over2})(n_{i,2j+1}-{1\over2})
+ J'_{\perp}\sum_{i,j}
(n_{i,2j+1}-{1\over2})(n_{i,2j+2}-{1\over2}) - h\sum_{i,j}c_{i,j}^{\dag}c_{i,j}
\label{hamiltonien}
\end{eqnarray}
using the two-dimensional
(2D) generalization of Jordan-Wigner transformation
where the phases $\Phi$, $\Phi_\perp$ and $\Phi'_\perp$ are readily 
derived.\cite{azzouz2}
The BMFT treatment consists, first and foremost, to approximate 
the sum of these phases on
each plaquette by $\pi$.
Then, quartic terms corresponding to Ising interactions are decoupled
by introducing the bond parameters $Q=\langle c_{i,j} c_{i+1,j}^{\dag}\rangle$,
$P=\langle c_{i,2j} c_{i,2j+1}^{\dag}\rangle$ and 
$P'=\langle c_{i,2j+1} c_{i,2j+2}^{\dag}\rangle$ that are calculated by minimising the 
mean-field free energy.
%
%\begin{eqnarray}
%F=JQ^2 + J_\perp P^2/2 + J'_{\perp}P'^2/2 + h/2\
%-{1\over{2\beta}}\int{{{\rm d}^2k}\over{(2\pi)^2}}\sum_{p=\pm}
%\ln\bigl(1+{\rm e}^{-\beta E_{p}({\bf k})}\bigr)
%\label{free}
%\end{eqnarray}
%
The dispersion relation is found to be
\begin{eqnarray}
E_{\pm}({\bf k})=\pm{1\over2}\{4J_1^2\sin^2k_x
+ 2J_{\perp1} J_{\perp2}\cos^2k_y
+J_{\perp1}^2
+ J_{\perp2}^2 + 4J_1(J_{\perp1}-J_{\perp2})\sin k_x\sin k_y\}^{1/2}-h
\label{dispersion}
\end{eqnarray}
with $J_1=J(1+2Q)$, $J_{\perp1}=J_\perp(1+2P)$ and $J_{\perp2}=J'_\perp(1+2P')$,
%
%\begin{eqnarray}
%Q=\int {d^2k\over(2\pi)^2} [f_-({\bf k})-f_+({\bf k})]
%[2J_1\sin^2k_x + (J_{\perp1} - J_{\perp2})\sin k_x\sin k_y]/4E_+({\bf k}),
%\label{moyen1}
%\end{eqnarray}
%
%\begin{eqnarray}
%P=\int {d^2k\over(2\pi)^2} &&[f_-({\bf k})-f_+({\bf k})]
%[J_{\perp1} +  J_{\perp2}\cos(2k_y) + 2J_1\sin k_x\sin k_y]/4E_+({\bf k}),
%\label{moyen2}
%\end{eqnarray}
%
%\begin{eqnarray}
%P'=\int {d^2k\over(2\pi)^2} &&[f_-({\bf k})-f_+({\bf k})]
%[J_{\perp2} +  J_{\perp1}\cos(2k_y) - 2J_1\sin k_x\sin k_y]/4E_+({\bf k}),
%\label{moyen3}
%\end{eqnarray}
%

In this work, elementary spin excitations are described in a fermionic like picture. 
Consequently, the number of fermions in excitations that contribute
to thermodynamical functions is found to be conserved just as in normal Fermi
systems. This result indicates that only the $S^z=0$ component of the triplet excited state
is relevant to thermodynamics, and is similar to what has been already verified 
both experimentally and theoretically in 
the case of spin-Peierls materials.\cite{azzouz3,azzouz4,kuo,mario} 
For example, this property is reflected by the fact that the magnetic Zeeman term 
is missing in the magnetic field dependence of the zero inter-ladder coupling 
energy gap $E_g(T,h)=|J_{\perp1}-J_{\perp2}|/2$.
The simple physical reason behind this effect is that thermal fluctuations
cannot create spin configurations with $S^z=\pm1$ because
up and down spins are equally distributed. This also remains 
true at zero temperature because $\langle S^z\rangle=0$ due to quantum fluctuations.
The energy gap, $E_g(T,h)$, of the single ladder ($J_\perp'=0$)
reported in Fig.\ \ref{susceptibility} as a function of temperature 
never vanishes, but instead decreases with increasing temperature.
In other words,
the effect of the gap is washed out by thermal fluctuations
when the temperature becomes larger than a characteristic value of the order 
of the zero temperature energy gap. This is in perfect agreement with the 
absence of a phase 
transition in the two real ladder materials we mentioned above.\cite{phase}
Also, from Fig.\ \ref{susceptibility}, it is seen that $E_g(T,h)$ remains 
unaffected by the magnetic field  
at low temperature. As we explained before, this result is reminiscent of 
the fact that only the $S^z=0$ component of the triplet excited state will 
contribute to thermodynamical quantities, and is a consequence of the absence of the 
magnetic Zeeman term in $E_g(T,h)$. The magnetic field dependence
is implicitly included in the parameters $Q$, $P$ and $P'$ and vanishes
for $T\to0$ as can be noticed in Fig.\ \ref{susceptibility}.
We should mention here that in the single ladder limit ($J_\perp'=0$),
the gap at $k_x=0$ is the same as that 
at $k_x=\pi$. This contrasts with exact diagonalization result\cite{barnes}
where the gap at $k_x=0$ is twice the gap at $k=\pi$. Our mean-field gap finds 
however a compromise by being half way between these gaps. 

As for inter-ladder coupling,
from Eq.\ (\ref{dispersion}), one can distinguish three regimes:
{\it i}) $J_\perp'=0$ where the only transverse modes are $k_y=0,\ \pi$. The third term
under the square root in the dispersion relation, that becomes equivalent 
to that of the ladder system, vanishes.
{\it ii}) $J_\perp'=J_\perp$, if $J_\perp\sim J$ then the AF order parameter
has to be included.\cite{azzouz2} In the controversial limit $J_\perp\ll J$,
the ground state is either ordered antiferromagnetically\cite{azzouz2} 
or is a gapless spin liquid state. The later is well described
by Eq.\ (\ref{dispersion}) where $(J_{\perp1}-J_{\perp2})\sin k_x\sin k_y=0$
since $J_{\perp1}=J_{\perp2}$.
{\it iii}) $J_\perp'\ll J_\perp$, this is an interesting regime. The dispersion
relation is found to present nodes at incommensurate wavevectors.
The third term under the square root,
which is absent in cases {\it i} and {\it ii}, is responsible for a such behavior.
For $J_\perp'=0.01$ for example,
the zeros of the spectrum occur at $(k_x,k_y)=(3.661212,\pm\pi/2)$,
$(k_x,k_y)=(5.763566,\pm\pi/2)$ and their images with respect to the symmetry
centre $(\pi,\pi)$ as shown in figure\ \ref{suscepcouple}.

Now, we calculate the momentum transfer- and frequency-dependent response function, 
$\chi({\bf q},\omega)$, by evaluating the space and time 
Fourier transform of the correlation function 
$\langle T(S_{i',j'}^z(\tau)S_{i,j}^z(\tau'))\rangle$. We obtain:
\begin{eqnarray}
\chi({{\bf q},\omega})={1\over8}\sum_{\bf k}\biggl\{\{{{f_+({\bf k}+{\bf q})-f_+({\bf k})}
\over{E_+({\bf k}+{\bf q})-E_+({\bf k})-\omega}} + (+\to-)\bigr\}\theta_+ + 
\bigl\{{{f_-({\bf k}+{\bf q})-f_+({\bf k})}
\over{E_-({\bf k}+{\bf q})-E_+({\bf k})-\omega}} + (+\to-)\bigr\}\theta_-
\biggr\}
\label{rfunction}
\end{eqnarray}
where $f_\pm$ is the Fermi factor and
$\theta_{\pm}= 1\pm\cos(\alpha_k-\alpha_{k+q})$ with $\alpha_k$, satisfying
$\tan\alpha_{k}={{2J_1\sin k_x + (J_{\perp1}-J_{\perp2})\sin k_y}
\over{(J_{\perp1}+J_{\perp2})\cos k_y}}$, is the phase of the coherence factors
required in the diagonalization of the mean-field Hamiltonian.
The static and uniform spin susceptibility, $\chi'(T,h)$, is obtained by evaluating
${\rm lim}_{{\bf q}\to0,\omega\to0}\chi({\bf q},\omega)$. 
This yields the following expression:
$\chi'(T,h)={\beta\over4}\sum_{{\bf k},p=\pm}(1+e^{\beta E_p{\bf k}})^{-2}
e^{-\beta E_{p}({\bf k})}.$
%
%For the imaginary part of $\chi({\bf q},\omega)$ that is relevant to 
%the calculation of the relaxation rate ${T_1}^{-1}$, it follows from 
%(\ref{rfunction}) that:
%
%\begin{eqnarray}
%\chi''({\bf q},\omega)=&&{\pi\over8}\sum_{\bf k}\bigl\{[f_+({\bf k}+{\bf q})
%-f_+({\bf k})]
%\delta[\omega-E_+({\bf k}+{\bf q}) + E_+({\bf k})] \cr
%&&\ \ \ \ \ + [f_-({\bf k}+{\bf q})-f_-({\bf k})]\delta[\omega-E_-({\bf k}+{\bf q})
%+ E_-({\bf k})]
%\bigr\}\theta_+({\bf k},{\bf q})\cr
%&&+{\pi\over8}\sum_{\bf k}\bigl\{[f_-({\bf k}+{\bf q})-f_+({\bf k})]
%\delta[\omega-E_-({\bf k}+{\bf q}) + E_+({\bf k})]\cr
%&&\ \ \ \ \ + [f_+({\bf k}+{\bf q})-f_-({\bf k})]\delta[\omega-E_+({\bf k}+{\bf q})
%+ E_-({\bf k})]
%\bigr\}\theta_-({\bf k},{\bf q}).
%\label{imaginary}
%\end{eqnarray}
%
The relaxation rate is calculated using Moriya's\cite{moriya} expression 
${1\over T_1}=2\gamma_N^2(g\mu_B)^{-2}T\omega^{-1}\int {{\rm d}{\bf q}\over(2\pi)^2}
{|A({\bf q})|^2\chi''({\bf q},\omega)}\ (\omega\to0)
$
where $\chi''({\bf q},\omega)$ is the imaginary part of $\chi({\bf q},\omega)$ and
$A({\bf q})$ is the nuclear hyperfine form factor that will be determined
by fiting experimental data. $\mu_B$, $g$ and $\gamma_N$ 
are respectively, the Bohr magneton, the Land\'e and gyromagnetic factors.
%$T_1^{-1}$ is calculated using 
%
%$${1\over{x+i\eta}}={x\over{x^2+\eta^2}}-
%i{\eta\over{x^2+\eta^2}},\ \ \ \ \ \eta\to0$$
%
%as a representation for the $\delta$ function which yields 
%$\delta(x)=\eta/(\pi\eta^2+\pi x^2)$.
In Fig.\ \ref{susceptibility}, we display 
$\chi'(T,h)$ as a function of temperature for several values of 
the magnetic field for the single ladder limit.
From this figure, one can notice that the magnetic field manifests itself
only at low temperature contrary to the energy gap that is rather affected at high
temperature. The dominant contribution to $\chi'(T)$ when $T\ll E_g$
that is given by $\cosh(\beta h)e^{-\beta E_g}$ ($E_g$ is field independent
for $T\sim0$) explains such a behavior.
In order to confirm these predictions in experiment, we suggest 
to carry on a measurement of the spin susceptibility
as a function of temperature and magnetic field on a ladder material.
Note finally that a sizable effect of the magnetic 
field begins at $h\sim 0.1E_g(T=0)$.
For weakly coupled ladders, as shown in Fig.\ \ref{suscepcouple}
for $J_\perp'=0.01\ {\rm and}\ J_\perp=J=1$, the energy spectrum is found to show nodes
(zero energies) at incommensurate wavevectors, $(k_x,k_y)=(3.661212,\pm\pi/2)$,
$(k_x,k_y)=(5.763566,\pm\pi/2)$ and their images.
As a result of these nodes, $\chi'(T)\sim T$ at low temperature.
Note that in a work reported by Gopalan {\it et al.},\cite{gopalan}
the gap vanishes only for $J_\perp'/J_\perp\ge0.25$ at commensurate
wavevectors. However, as their mean-field theory favors the opening 
of an energy gap, the question of the right threshold value 
(wether it is $0$ or $0.25$) will remain open. An argument in favor of our
result is given by the fact that nodes occur at incommensurate wavevectors
as a consequence of coupling anisotropy.

%
%\section{Application to experiment}
%

Let us compare with experiment. First we would like to address 
the temperature dependence of the energy gap. In Fig.\ \ref{susceptibility}, 
one notes that the gap decreases with increasing temperature. 
So, contrary to previous 
investigations,\cite{azuma,ishida} where the zero temperature 
value of the gap is used for all temperatures,
one should take into account the decrease of the gap.
The data of Johnston {\it et al.}\cite{Johnston} concerning the uniform
magnetic susceptibility
for the (VO)$_2$P$_2$O$_7$ compound are fitted in Fig.\ \ref{NMR} 
using the single
ladder limit. These data have already been fitted by Barnes and Riera\cite{barnes} 
using exact diagonalization with $J=7.82$ meV and $J_\perp=7.76$ meV. We have noted
that our calculations for the single ladder are more accurate in this case than for 
SrCu$_2$O$_3$ (we do not report the fit for SrCu$_2$O$_3$ because of lack of space).
Concerning the NMR relaxation rate, we plot in Fig.\ \ref{NMR}
both Azuma {\it et al.}'s experimental results of the SrCu$_2$O$_3$ compound
and the best fit obtained
using the same coupling constants as for the susceptibility.
This suggests, as an answer to the problem rose in 
Dagotto {\it et al.}'s work,{\cite{dagotto}} that both $\chi'$ and $1/T_1$ 
can be modeled using the same set of parameters of the Heisenberg single ladder.
For the magnetic field, a sizable effect in this material is expected
for fields $h\sim0.1E_g(T=0)\sim100$ T that are not 
accessible in experiment. One can instead use the (VO)$_2$P$_2$O$_7$ material 
since the energy gap $E_g\sim 50$ K
is much smaller.\cite{barnes,eccleston} This yields $h\sim10$ T.

The departure from experimental data at high temperature may be attributed to 
the fact that our spectrum is the same for either $k_\perp=0$ or $k_\perp=\pi$,
and gaps at $k=0$ and $k=\pi$ are equal. As stressed before,
the gap at $k=0$ is twice that at $k=\pi$.\cite{barnes}
Another point is that since energy gaps decrease with increasing temperature, 
contributions from the continuum of the spectrum above the gap at $k=0$ could become
not negligible already for temperatures $T\sim E_g(T=0)$.\cite{david} 
This is obvious for (VO)$_2$P$_2$O$_7$ where deviations start 
at around $T\sim50$ K. However, it is less clear for SrCu$_2$O$_3$ where
deviations are found to occur at $T\sim250$ K which is almost half 
the gap ($E_g\approx468$ K).
From these above remarques, we conclude that our theory can be applied to explain
experiments at low temperature ($T\le E_g(T=0)$).

%
%\section{Conclusion}
%

In conclusion, we have used the bond-mean-field theory to investigate 
AF Heisenberg ladder spin systems. The uniform spin susceptibility and the NMR 
spin-relaxation rate are calculated as a function of temperature and 
uniform magnetic field. 
For a single ladder, the energy gap
remains unaffected by the magnetic field at low temperature
in agreement with the fact that the relevant elementary spin excitations 
to thermodynamics (as seen here for $\chi'$ and $T_1^{-1}$) 
conserve the total $S^z=0$ component. Our results successfully 
compare to experiments on 
SrCu$_2$O$_3$ and (VO)$_2$P$_2$O$_7$ compounds for $T\le E_g$. Several 
predictions concerning the magnetic field effect that can be verified in simple
experiments are also made.
Inter-ladder coupling is found to drastically change the behavior of the uniform 
susceptibility, $\chi'$, at very low temperature. For $J_\perp'=0.01J_\perp$ 
for example, $\chi'$ becomes linear in $T$
as a consequence of the appearance of nodes in the spectrum at 
incommensurate wavevectors, $(k_x,k_y)=(3.661212,\pm\pi/2)$,
$(k_x,k_y)=(5.763566,\pm\pi/2)$ and their images.
It is worthy of note to mention, in the end of this work, that Gopalan 
{\it et al.}\cite{gopalan} suggested that the inter-ladder coupling in SrCu$_2$O$_3$
is ferromagnetic. They based their argument on the fact that 
two Cu ions on adjacent ladders are connected through an O site by $90^o$
bonds. A calculation in this case of inter-ladder coupling
will be reported in the near future.

\acknowledgements
We would like to thank A.-M. Tremblay and S. Moukouri 
for reading the manuscript and for helpful
comments. This work was supported by the Natural Sciences and 
Engineering Research Council of Canada (NSERC) and the Fonds pour 
la formation de chercheurs et l'aide \`a la recherche from the Government of 
Qu\'ebec (FCAR). 
\begin{figure}
\caption{(a) The susceptibility is drawn for $J_\perp'=0$ 
as a function of temperature for several values of $h$. In the inset, $E_g$ is 
drawn as a function of temperature for $h=0$.}
\label{susceptibility}
\end{figure}
\begin{figure}
\caption{The dispersion relation for weakly coupled ladders is ploted
as a function of $k_x$ and $k_y$; $J_\perp'=0.01J_\perp$.}
\label{suscepcouple}
\end{figure}
\begin{figure}
\caption{In (a), the fit to Johnston {\it et al.} susceptibility data
is reported for (VO)$_2$P$_2$O$_7$. The fit to Dagotto {\it et al.}'s 
experimental results for $1/T_1$ is shown in (b) in the case of SrCuO$_3$.
Data are drawn in crosses.}
\label{NMR}
\end{figure}


\begin{references}
\bibitem{phase} If the gap should vanish, the situation would be similar to spin-Peierls
materials where it vanishes at $T_{SP}$.
\end{references}
\end{document}